\pdfoutput=1 
\documentclass[a4paper,12pt]{article}
\usepackage{bxpapersize}
\usepackage{graphicx}
\usepackage{times}
\usepackage{latexsym}
\usepackage{algorithm}
\usepackage{algorithmic}
\usepackage{amsmath}
\usepackage{amsthm}
\usepackage{ascmac}
\usepackage{mathrsfs}
\usepackage{multirow}
\usepackage{amssymb}
\usepackage{color}
\usepackage{framed}
\usepackage{fancybox}
\usepackage{wrapfig}

\usepackage{amsmath}
\usepackage{amsfonts}
\usepackage{bbm} 

\newcommand{\bbE}{\mathbb{E}}

\newcommand{\bbP}{\mathbb{P}}

\newcommand{\bbR}{\mathbb{R}}
\newcommand{\bbS}{\mathbb{S}}
\newcommand{\bbV}{\mathbb{V}}

\newcommand{\cH}{\mathcal{H}}

\newcommand{\cX}{\mathcal{X}}

\newcommand{\nn}{\nonumber}

\def\str1#1#2{\left[ \begin{array}{c} #1 \\ #2 \end{array}\right]}

\newtheorem{theorem}{Theorem}[section]

\newtheorem{lemma}[theorem]{Lemma}
\newtheorem{corollary}[theorem]{Corollary}
\newtheorem{definition}[theorem]{Definition}

\makeindex
\allowdisplaybreaks[1]
\title{On Learning from Ghost Imaging without Imaging}
\author{Issei Sato \\ The University of Tokyo / RIKEN\\sato@k.u-tokyo.ac.jp}
\date{}
\begin{document}
\maketitle
\begin{abstract}
Computational ghost imaging is an imaging technique in which an object is imaged from light collected using a single-pixel detector with no spatial resolution.
Recently, ghost cytometry has been proposed for a high-speed cell-classification method that involves ghost imaging and machine learning in flow cytometry.
Ghost cytometry skips the reconstruction of cell images from signals and directly used signals for cell-classification because this reconstruction is what creates the bottleneck in high-speed analysis. 
In this paper, we provide a theoretical analysis for learning from ghost imaging without imaging. 
\end{abstract}
\section{Introduction}
Ghost imaging was first observed with entangled photon pairs and viewed as a  quantum phenomenon \cite{Pittman:PRA:1995}.
It acquires object information through the correlation calculations of the light-intensity fluctuations of two beams: object and reference \cite{Erkmen:AOP:10,Shapiro:QIP:2012}.
Computational ghost imaging is an imaging technique in which an object is imaged from illumination patterns and light collected  using a single-pixel detector with no spatial resolution \cite{Shapiro:PRA:2008,Katz:APL2009},
which simplifies the operations in comparison to conventional two-detector ghost imaging.
Photomultiplier tube (PMT) is often used as a single-pixel detector to collect the scattered light.
Using the detected signals and illumination patterns enables us to computationally reconstruct images.

Let $T(x,y)$ be the transmission function of an object.
An object is illuminated by a speckle field generated by passing a laser beam through an optical diffuser, which is a material that diffuses light to transmit light such as diffractive optical elements (DOEs).
A detector measures the total intensity, $G_m$, transmitted through the object given by
\begin{align}
G_{m}=\int I_{m}(x,y)T(x,y)dxdy,
\end{align}
where $I_{m}(x,y)$ is the $m$-th speckle field, which is also refereed as  the $m$-th  structured illumination pattern. 

We can reconstruct the transmission function expressed by
\begin{align}
\widetilde T(x,y)=\frac{1}{M}\sum_{m=1}^{M} (G_{m}-\langle G \rangle)I_{m}(x,y),~\text{where}~\langle G \rangle=\frac{1}{M}\sum_{m=1}^{M}G_{m}.
\end{align}
The transmission function $\widetilde T(x,y)$ indicates a reconstructed image of the target object.

\textit{Ghost cytometry} \cite{Ota:Science:2018} has been proposed as a high-speed cell-classification method and involves ghost imaging and machine learning in \textit{ flow cytometry}.
Flow cytometry is a technique to measure the characteristics of a population of particles (cell, bacteria etc.) such as cell size, cell count, cell morphology (shape and structure), and cell cycle phase at high speed .
\textit{Cyto-} and \textit{-metry} mean cell and measure, respectively.
With flow cytometry, we can measure the information of a single cell. 
A sample including cells, e.g., blood cells , is injected into a flow cytometer, which
is composed of three systems: flow/fluid, optical, and electric.
It detects scattered light and the fluorescence of cells.
From the detected scattered light and fluorescence signals, we can obtain information on the relative size and internal structure of a cell and the cell membrane, cytoplasm, various antigens present in the nucleus, and quantities of nucleic acids.

Computational ghost imaging is well known as an imaging method.
However, a breakthrough occurred with ghost cytometry in which the reconstruction of cell images from raw signals  $\{G_{m}\}_{m=1}^{M}$  can be skipped because this reconstruction is what causes the bottleneck in high-speed analysis. 
Ghost cytometry directly uses raw signals to classify cells.
Also, computational ghost imaging uses randomly generated multiple illumination patterns to reconstruct an image.  
In ghost cytometry, cells pass through a randomly allocated illumination pattern and the signals are detected in time series using a single pixel detector. 
That is, we do not need to switch the illumination pattern to obtain the fluorescence-intensity features extracted from multiple illumination patterns, which differs from ghost imaging.

In ghost cytometry,  morphologically similar but different types of cells are classified.
According to a subsequent study \cite{Ota:arXiv2019}, it was not possible to classify these different types of cells by using the two commonly available features, namely, the intensity of fluorescence and  forward scattering intensity, of a commercialized flow cytometer, \textit{JSAN}.
However, an image-based support vector machine (SVM), which is learned from $28 \times 28$ pixels images obtained using a commercialized image flow cytometer, \textit{ImageStreamX}, achieved a test AUC of $0.967$.
This result is the same as that of image-free-based SVM in ghost cytometry.
Thus, we consider that ghost features  capture some morphological information in raw signals.

In this paper, we provide a theoretical analysis for learning from ghost imaging without imaging both general ghost imaging and ghost cytometry settings. 
We show that ghost features approximate the radial basis function (RBF) kernel between object images by using signals without imaging.
That is, 
\begin{align}
\kappa_{\mathrm{RBF}}(X,Y)  \approx \kappa_{\mathrm{RBF}}(G(X),G(Y)),
\end{align}
where $\kappa_{\mathrm{RBF}}(X,Y)$ is the RBF kernel between image objects $X$ and $Y$ and $\kappa_{\mathrm{RBF}}(G(X),G(Y))$ is the RBF kernel between signals $G(X)$ and $G(Y)$  in ghost cytometry.

\section{Related work}\label{sec:related_work}
Recently, \textit{optical machine learning}, which is a fusion field of optics and machine learning,  is receiving increased attention.
Lin et al \cite{Lin:Science:2018} developed all-optical diffractive deep neural network that is a physical mechanism by using 3D-printer.
They classified handwritten digits and fashion images.
This framework can perform in all-optical image analysis, feature detection, and object classification.
Ghost cytometry is regarded as a study of this line in which random projection is used as a machine learning method and implemented by using diffractive optical elements (DOEs) to extract features from a cell object. The feature extraction, which is essential yet creates a bottleneck in image processing, performs at the speed of light by using optical elements.
The relationship between optics and neural networks goes back to the 1980s  \cite{WagnerPsaltis:AO1987, Caulfield:IEEE1989}.
In recent years, this trend may re-emerge with new technologies on both sides.
Compared to an all-optical system, ghost cytometry uses a hybrid system in which the feature extraction is implemented on DOEs and PMT, and the classifier is implemented on a field programmable gate array (FPGA). This hybrid system seems more flexible than an all-optical system because fine tuning is easier in FPGA than in DOEs.

The algorithm to create ghost features is regarded as a kind of random projection method.
Dimensionality reduction based on random projection, which has been well studied in machine learning, is built on the idea that high-dimensional data lie in fact in a lower-dimensional subspace.
The breakthrough occurred with the Johnson-Lindenstrauss lemma \cite{JL1984} in $1984$. 
It states that there exists a mapping from a high-dimensional space into a lower-dimensional space that can preserve the pairwise Euclidean distances between data points up to a bounded relative error.
Dasgupta and Gupta \cite{DG2003} provided this lemma by using elementary probabilistic techniques.
Achlioptas \cite{Achlioptas:JCSS2003} relaxed the Gaussian distribution to a discrete random distribution with zero mean and variance one, i.e., a sparse random projection using the random matrix $R$ with i.i.d entries in
$\{\sqrt{s},0,-\sqrt{s}\}$ with probabilities $\{\frac{1}{2s},1-\frac{1}{2s},\frac{1}{s}\}$ for $s=1$ and $s=3$. 
Li et al. \cite{Li:KDD2006} improved their work considering $s>3$.
Matousek \cite{Matousek08} generalized such results to sub-Gaussian random variables. 

An interesting intersection of neural networks and random projections is \textit{neuron-friendly random projection} \cite{ArriagaVempala:FOCS1999}.
They proposed random projection to learn robust concept classes from a few examples with a biologically plausible neuronal mechanism called \textit{neuronal random projection}.
Neuronal random projection has a a random matrix whose entries are chosen independently from standard normal distribution or uniform distribution over $\pm 1$.
Robust concept learning is closely related to large margin classifiers.
Shi et al. \cite{Shi:ICML2012} analyzed margin preservation for binary classification problems where they showed results for random Gaussian projection matrix.
Although they only showed results for the random Gaussian matrix, similar bounds seems to be achieved for a subGaussian distribution.

As described in Section \ref{sec:analysis}, we use a Bernoulli random variable for the random projection matrix.
Although a Bernoulli random variable can be regarded as a subGaussian random variable, 
we cannot use this property to obtain tight bounds in our theory when the probability of coin toss is biased. Moreover, projection matrices are not independent. Therefore, we need to devise an analysis of ghost features in this paper that is different from existing work.

\section{Analysis}\label{sec:analysis}
Ghost features are regarded as a type of random projection; thus, we analyze them in terms of random projections.
We first describe our motivation and then define several terms and show lemmas to derive the main results: Theorems \ref{theorem:JL:L2} and \ref{theorem:JL:gc}.
The details of the proofs are provided in the supplementary material.

\subsection{Motivations and preliminary theorems}\label{sec:motivation}
In the analysis of random projection , i.i.d entries of a random matrix are generally assumed to be subGaussian \cite{Matousek08}.
\begin{definition}[$\sigma^2$-subGaussian]
A random variable $Z\in\bbR$ is said to be $\sigma^2$-subGaussian if  there exists $\sigma>0$ such that its moment generating function satisfies
\begin{align}
\forall \lambda \in\bbR,~\bbE[\exp(\lambda (Z-\bbE[Z]))]\leq \exp\left(\frac{1}{2}\sigma^2 \lambda^2\right).
\end{align}
The constant $\sigma^2$ is called a proxy variance.
\end{definition}
\begin{definition}[Optimal proxy variance \cite{BuldyginarMoskvichova:TPMS2013, Jin:arXiv2019}]
The smallest proxy variance is called the optimal proxy variance and is denoted $\sigma^2_{\mathrm{opt}}(Z)$, or simply $\sigma^2_{\mathrm{opt}}$. 
The variance $\bbV[Z]$ always provides a lower-bound on the optimal proxy variance.
When $\bbV[Z]=\sigma^2_{\mathrm{opt}}(Z)$, $Z$ is said to be strictly subGaussian.
\end{definition}

Let $b\in\{1,0\}$ be a Bernoulli random variable with parameter $q$ that is the occurrence probability of one.
Structured illumination patterns can be formulated as Bernoulli random matrices.
In ghost imaging and cytometry, one laser is divided into multiple ones using DOEs; thus, structural illumination patterns need to be sparse in order to improve the signal-to-noise (S/N) ratio. That is, $q$ needs to be small, which is problematic in analyzing ghost features.

The optimal proxy variance of a Bernoulli random variable has the form $\displaystyle \sigma^2_{\mathrm{opt}}(b)=\frac{\frac{1}{2}-q}{\log (\frac{1}{q}-1)}$.
The Bernoulli random variable is strictly subGaussian if and only if $q=\frac{1}{2}$.
However, the optimal proxy variance $\sigma^2_{\mathrm{opt}}(b)$ is larger than  variance $\bbV[b]=q(1-q)$ when $q$ is small, which means that the exponential inequality is too loose when we use the subGaussian property of a Bernoulli random variable. 
Moreover, projected features are not independent in the sense that they share projected matrices to obtain projected features. 
%
%
%
%
To overcome these problems. we use  two theorems: Theorems \ref{theo:bernstein_ineq} and \ref{theo:dependent_cov}.

Theorem \ref{theo:bernstein_ineq} is referred as the Bernstein inequality \cite{Bernstein:GPH1946} and has different forms; we consider the following one.
\begin{theorem}[Bernstein inequality]\label{theo:bernstein_ineq}
Let $Z$  be a random variable satisfying the Bernstein condition
\begin{align}
\bbE[|Z-\bbE[Z]|^k]\leq \frac{1}{2}k!\sigma^2C^{k-2}~(k=3,4,\ldots).
\end{align}
Then, for all $|\lambda| <1/C$,
\begin{align}
\bbE[\exp(\lambda Z)]\leq \exp\left(\lambda\bbE[Z]+\frac{\lambda^2\bbV[Z]}{2(1-|\lambda| C)}\right),
\end{align}
and the concentration inequality
\begin{align}
\bbP[|Z-\bbE[Z]|\geq \epsilon]\leq 2\exp\left(-\frac{\epsilon^2}{2(\bbV[Z]+C\epsilon)}\right)~\text{for all}~\epsilon\geq0.
\end{align}
\end{theorem}
One sufficient condition for the Bernstein condition to hold is that $Z$ be bounded;
in particular, if $|Z-\bbE[Z]| \leq C$, then it is straightforward that the Bernstein condition hold.


The following theorem is known on non-negatively associated random variables \cite{Dewan:JNS1999, Newman:CMP1980}.
\begin{theorem}\label{theo:dependent_cov}
Let $\{Z_i\}_{i=1}^{n}$ be non-negatively associated random variables bounded by a
constant $C$ and $\mathrm{Cov}(Z_i,Z_j)$ be the covariance of $Z_i$ and $Z_j$. Then for any $\lambda>0$,
\begin{align}
\left| \bbE\left[\exp\left(\lambda\sum_{i=1}^{n}Z_i\right)\right] - \prod_{i=1}^{n}\bbE[\exp\left(\lambda Z_i\right)] \right|\leq \lambda^2\exp(n\lambda C)\sum_{1\leq i < j\leq n}\mathrm{Cov}(Z_i,Z_j).
\end{align}
\end{theorem}

\subsection{Analysis of Ghost Features in Ghost Imaging}\label{sec:analysis}
Let $B_m$ be an $H \times W$ random binary matrix where $m\in\{1,2,\ldots,M\}$.
The $(i,j)$-th element, $B_m(i,j)$, indicates the $m$-th speckle field  $I_{m}(x_i,y_j)$.
We construct $B_m(i,j)$ by using
\begin{align}
B_m(i,j)=
  \begin{cases}
    1~\text{with probability $q$},\\
    0~\text{with probability $1-q$},
  \end{cases}
\end{align}
where $q\in(0,1)$ is a parameter.

Denote an $H \times W$ matrix representing an object as $X$ , i.e.,
the $(i,j)$-th element, $X(i,j)$, indicates the value of a transmission function of an object, given by $X(i,j)=T(x_i,y_j)$.
Therefore, we reformulate $G_m$ measured using a detector, given by
\begin{align}
G_{m}(X)= \sum_{i=1}^{H}\sum_{j=1}^{W}B_{m}(i,j)X(i,j).\label{def:gi:features}
\end{align}
We can reconstruct 
\begin{align}
\widetilde X(i,j)=\frac{1}{M}\sum_{m=1}^{M} (G_{m}(X)-\langle G(X) \rangle)B_{m}(i,j),
\end{align}
where $\displaystyle \langle G(X) \rangle=\frac{1}{M}\sum_{m=1}^{M}G_{m}(X)$.
However, we consider learning from ghost imaging without image reconstruction.
We call $\{G_m(X)\}_{m=1}^{M}$ \textit{ghost features} of object $X$.

We define the $m$-dimensional vector function expressed by
\begin{align}
G(X)&=(G_1(X), G_2(X), \ldots, G_M(X))^{\top},\\
g(X)&=(g_1(X), g_2(X), \ldots, g_M(X))^{\top},~
g_m(X)=G_{m}(X)-\langle G(X) \rangle,
\end{align}
where $\top$ is a transpose of a vector and matrix.

\begin{definition}[L$2$ norm and Frobenius norm]
Denote the L$2$ norm of vector $g$ as $\|g\|_2$  and  Frobenius norm of matrix $X$ as $\| X \|_{\mathrm{F}}$.
\end{definition}

\begin{definition}[Summation of matrix elements]
Let the summation of matrix elements be
\begin{align}
\bbS[X]=\sum_{i=1}^{K}\sum_{j=1}^{K}X(i,j).
\end{align}
\end{definition}

\begin{definition}[Remainder and Quotient]
Denote the remainder and quotient upon division of $A$ by $B$ as $[A\%B]$ and $\lfloor A/B\rfloor$, respectively.
\end{definition}
We define two functions used in the following lemma and theorem.
We will explain their meanings in Remark\,1 below.
\begin{definition}\label{def:gamma_lambda}
Denote $i'=[k'\%H]$, $j'=\lfloor k'/H\rfloor$ and define
\begin{align}
\Gamma_q(X)&\overset{\mathrm{def}}{=}\frac{(1-2q)^2}{q(1-q)}\sum_{j=1}^{W}\sum_{i=1}^{H}X(i,j)^4 +4\sum_{j=1}^{W}\sum_{i=1}^{H}\sum_{k'>(j-1)H+i}^{WH}(X(i,j)X(i',j'))^2,\\
\Lambda_q(X)&\overset{\mathrm{def}}{=}\max\left\{\max_{(i,j)\not=(i',j')}2\left|\frac{1-q}{q}X(i,j)X(i',j')\right|,~\max_{(i,j)}\left|\frac{1-2q}{q}X(i,j)^2\right|\right\}.
\end{align}
\end{definition}

The feature vector $g(\cdot)$ has the linear property.
\begin{lemma}[Linearity]\label{prop:gf:linear}
Let $X$ and $Y$ be $N \times N$ real matrices.
\begin{align}
g_m(X-Y)=g_m(X)-g_m(Y).
\end{align}

\end{lemma}

\noindent Since we consider the two parts of $g_m(X)$ in the main result below,
\begin{align}
g_m(X)=G_m(X)-\langle G(X) \rangle=\underbrace{G_m(X)-q\bbS[X]}_{\displaystyle \text{Part I}} + \underbrace{q\bbS[X] - \langle G(X) \rangle }_{\displaystyle \text{Part II}}. 
\end{align}
we show the exponential  inequality of the two parts by using Theorems \ref{theo:bernstein_ineq} and \ref{theo:dependent_cov} as follows.
\begin{lemma}[Exponential inequality]\label{lemma:gi:exp_ineq:part_1}
For any $\displaystyle \frac{|t|}{M} < \frac{1}{\Lambda_q(X)}$,
\begin{align}
&\bbE\left[\exp\left(\frac{t}{q(1-q)M}\sum_{m=1}^{M} (G_m(X)-q\bbS[X])^2\right)\right]
\leq \exp\left(t\|X\|_{\mathrm{F}}^2+\frac{\Gamma_q(X)}{2M\left(1-\frac{\Lambda_q(X)}{M} |t|\right)}t^2\right),\\
&\bbE\left[\exp\left(\frac{t}{q(1-q)}(q\bbS[X]-\langle G(X) \rangle)^2\right)\right]
\leq \exp\left(\frac{t}{M}\|X\|_{\mathrm{F}}^2+\frac{\Gamma_q(X)}{2M^3\left(1-\frac{\Lambda_q(X)}{M^2}|t| \right)}t^2\right).
\end{align}
\end{lemma}
Since $\{g_m(X)\}_{m=1}^{M}$ share $\langle G(X) \rangle$, they are not independent given $X$, which is the difference from existing random projections.
By using Lemma \ref{lemma:gi:exp_ineq:part_1} and H\"{O}lder's inequality, we have the following theorem.
\begin{theorem}\label{theorem:JL:L2}
Let $X$ and $Y$ be $H \times W$ real matrices.
For all $\epsilon$, with probability at least $1-\delta$,
\begin{align}
\left(1-\frac{1}{M}-\epsilon\right)\| X-Y \|_{\mathrm{F}}^2 \leq \frac{1}{Mq(1-q)}\|g(X)-g(Y)\|_2^2\leq \left(1-\frac{1}{M}+\epsilon \right) \| X-Y \|_{\mathrm{F}}^2,
\end{align}
where
\begin{align}
\delta=2\exp\left(-\frac{\epsilon^2M}{2\left(\left(1+\frac{2}{M^2}\right)\frac{\Gamma_q(X-Y)}{\|X-Y\|_{\mathrm{F}}^4}+\frac{\Lambda_q(X-Y)}{\|X-Y\|_{\mathrm{F}}^2}\epsilon\right)}\right).
\end{align}
\end{theorem}

\textbf{Remark\,1.}
The number of illumination patterns, $M$, need to be larger when $\epsilon^2$ is smaller.
This is the same property of existing random projections as the Johnson-Lindenstrauss lemma. In our case, we need to increase $M$ according to $q$ because $\Gamma_q$ and $\Lambda_q$ increase when $q$ decreases, which is reasonable because obtaining information out of a sparse matrix requires increasing the number of the sparse matrix. In Definition \ref{def:gamma_lambda}, $\Gamma_q$ and $\Lambda_q$ consist of two parts: intensity and correlation. In $\Gamma_q(X-Y)$ and $\Lambda_q(X-Y)$, $(X(i,j)-Y(i,j))^2$ indicates the intensity and $(X(i,j)-Y(i,j))(X(i',j')-Y(i',j'))$ is the pixel-wise correlation of the difference image between $X$ and $Y$. The intensity of $X-Y$ is small when $X$ and $Y$ are similar objects, otherwise large. The correlation of $X-Y$ is small when $X-Y$ is sparse. When we use the maximum value of  $\Gamma_q(X-Y)$ and $\Lambda_q(X-Y)$ over the space of $X-Y$, Theorem \ref{theorem:JL:L2} is independent of objects. In fact, the element of $X$ is bounded and controllable to some extent.

\subsection{Analysis of Ghost Features in Ghost Cytometry}\label{sec:analysis}

In ghost imaging, multiple illumination patterns are independent, i.e., $\{B_m\}_{m=1}^{M}$ are independently and randomly generated.
Thus, the detected signals, i.e., ghost features $\{G_m\}_{m=1}^{M}$, do not share  illumination patterns $\{B_m\}_{m=1}^{M}$, i.e., $G_m$ is generated only from $B_m$.
In ghost cytometry, however, objects pass through a randomly allocated illumination pattern; thus, the detected features share illumination patterns as follows.

Let $B$ be $H\times M$ random binary masks where the $(i,j)$-th element, B(i,j), is constructed by
\begin{align}
B(i,j)=
  \begin{cases}
    1~\text{with probability $q$},\\
    0~\text{with probability $1-q$},
  \end{cases}
\end{align}
where $q\in[0,1]$ is a parameter.
The matrix $B$ is a illumination pattern in ghost cytometry.

The ghost feature for fluorescence object $X$ is formulated as
\begin{align}
G_m(X)=\sum_{i=1}^{H} \sum_{j=1}^{W} B(i,j+m-W) X(i,j),\label{def:gc:features}
\end{align}
where for simplicity of notation, if $j<0$ and $j>M$, $B(i,j)=0$.
The problem is that $G_1(X),G_2(X),\ldots,G_{M+W-1}(X)$ are highly correlated because they share the elements of $B$.
%
Ghost cytometry uses Ghost features $G_1(X),G_2(X),\ldots,G_{M+W-1}(X)$ to classify cell types.
We analyze ghost features obtained from Eq\,.\eqref{def:gc:features}.

It is worth noting the following.
Since the time for cells to pass through the structural illumination is several microseconds and the length of the structural illumination is several micrometers, it can be assumed that by fluid control, the cell does not rotate and passes through the center of the structural illumination.

Similar to Definition \ref{def:gamma_lambda}, we define the following two functions.
\begin{definition}\label{def:psi_phi}
Denote $i'=[k'\%H]$ and $m'=\lfloor k'/H\rfloor$ and define
\begin{align}
\Psi_q(X)
&\overset{\mathrm{def}}{=}\frac{(1-2q)^2}{q(1-q)}\sum_{i=1}^{H}\left(\sum_{j=1}^{W}X(i,j)^2\right)^2\nn\\
&\quad+4\sum_{i=1}^{H}\sum_{k'>(m-1)H+i}^{(m-1+W)H}\left(\sum_{j=1}^{W-(m'-m)}X(i,j)X(i',j+m'-m)\right)^2,\\
\Phi_q(X)&\overset{\mathrm{def}}{=}\max\left\{\max_{(i,m)\not=(i',m')}2\left|\frac{1-q}{q}\sum_{j=1}^{W-(m'-m)}X(i,j)X(i',j+m'-m)\right|,\right.\nn\\
&\hspace{2cm}\left.\max_{i}\left|\frac{(1-2q)}{q}\sum_{j=1}^{W}X(i,j)^2\right|\right\}.
\end{align}
\end{definition}
Compared to $\Gamma_q$ and $\Lambda_q$, $\Psi_q$ and $\Phi_q$ are a little complicated but their basic meanings are the same as those of $\Gamma_q$ and $\Lambda_q$ described in Remark\,1.

By using Theorems \ref{theo:bernstein_ineq} and \ref{theo:dependent_cov}, we obtain the following theorem.
\begin{theorem}\label{theorem:JL:gc}
Let $X$ and $Y$ be $H \times W$ real matrices.
For all $\epsilon$, with probability at least $1-\delta$,
\begin{align}
\frac{1}{q(1-q)(M+W-1)}\|G(X)-G(Y)\|_2^2\geq\left(1-\epsilon\right)\| X-Y \|_{\mathrm{F}}^2-\frac{q}{(1-q)}\bbS[ X-Y ]^2, \\
\frac{1}{q(1-q)(M+W-1)}\|G(X)-G(Y)\|_2^2 \leq \left(1+\epsilon \right) \| X-Y \|_{\mathrm{F}}^2+\frac{q}{(1-q)}\bbS[ X-Y ]^2,
\end{align}
where
\begin{align}
\delta=2\exp\left(-\frac{\epsilon^2M}{2\left(\frac{\Psi_q(X-Y)}{\|X-Y\|_{\mathrm{F}}^4}+\frac{\Phi_q(X-Y)}{\|X-Y\|_{\mathrm{F}}^2}\epsilon\right)}\right).
\label{eq:gc:delta}
\end{align}
\end{theorem}

\textbf{Remark\,2.}
The basic property of Theorem \ref{theorem:JL:gc} is the same as that described in Remark\,1 on Theorem \ref{theorem:JL:L2}. 
One of the differences between Theorems \ref{theorem:JL:L2} and \ref{theorem:JL:gc}  is that we evaluate $g$ or $G$ because In ghost cytometry, $G$ is directly used in classification.  The difference appears in the presence or absence of $\frac{q}{(1-q)}\bbS[X-Y]$. 
It is desirable that $q$ be small because one laser is divided into multiple ones using DOEs and the binary matrix needs to be sparse for the better signal-to-noise ratio. 
That is,  the term $\frac{q}{(1-q)}\bbS[X-Y]$ is small. 
Moreover, $\bbS[X-Y]$ is typically small.
When objects $X$ and $Y$ are similar,  $\bbS[X-Y]$ takes a small value.
When objects $X$ and $Y$ are dissimilar, $\bbS[X-Y]$ also takes a small value because the elements of matrix $X-Y$ takes positive and negative values and operation $\bbS$ is just a summation of the elements. When we use the formulation of $g$, the term $\bbS[X-Y]$ disappear and might improve the classification performance of ghost cytometry.

Since the following corollary holds, we also obtain the subExponential forms of Theorems \ref{theorem:JL:L2} and \ref{theorem:JL:gc}.
\begin{corollary}[subExponentiality]\label{col:subExponentiality}
Let $Z$  be a random variable satisfying Bernstein inequality, Theorems \ref{theo:bernstein_ineq}.
Then, for all $|\lambda| <1/(2C)$,
\begin{align}
\bbE[\exp(\lambda Z)]\leq \exp\left(\lambda\bbE[Z]+\lambda^2\bbV[Z]\right),\label{exp_inq:subExp}
\end{align}
and for all $|\epsilon| <\bbV[Z]/C$,
\begin{align}
\bbP[|Z-\bbE[Z]|\geq \epsilon]\leq 2\exp\left(-\frac{\epsilon^2}{4\bbV[Z]}\right)~\text{for all}~\epsilon\geq0.
\end{align}
\end{corollary}

\subsection{Discussion}\label{sec:discussion}
Theorems \ref{theorem:JL:L2} and \ref{theorem:JL:gc}  indicate that the RBF kernel function calculated using ghost features approximates the RBF kernel function using image objects, i.e.,
\begin{align}
\kappa_{\gamma}(X,Y)=\exp\left(-\gamma \|X-Y\|_{\mathrm{F}}^2 \right)\approx \kappa_{\beta}(g(X),g(Y))=\exp\left(-\beta \|g(X)-g(Y)\|_2^2\right),
\end{align}
where $\gamma \in (0,+\infty)$ and $\beta \in (0,+\infty)$ are kernel parameters.
Note that we can tune $\beta \in (0,+\infty)$  in stead of tuning $\gamma$ in the case of cross-validation.

The Frobenius norm is not rotation/shift-invariant to capture morphological information.
However, in the case of flow cytometry, we can obtain more representative objects by using  \textit{real} data augmentation from which we obtain augmented ghost features by injecting the object into the flow cytometer many times.

It is well known that a kernel function defines feature maps and vice versa.
Let $\cH$ be a Hilbert space.
A feature map $\phi: \cX \rightarrow \cH$ takes input $x \in \cX$ to infinite feature vectors $\phi(x) \in \cH$.
For every kernel $\kappa$, there exists Hilbert space $\cH$ and feature map $\phi:\cX\rightarrow\cH$ such that $k(x,x')=\langle \phi(x), \phi(x') \rangle$ where $\langle \cdot, \cdot \rangle$ is the inner product in the Hilbert Space.
That is, on the basis of kernel theory, when we focus on cell-image objects as input space, $\phi(X)$ indicates some of the features of cell-image object $X$.
The features on Hilbert space are black-box, but with SVM, we predict the label of a target object by using the labels of representative objects, called \textit{support vectors}, similar to the target object in terms of the Frobenius norm.
Thus, the representative objects may have specific morphological features for prediction.
That is, analyzing the representative cells may lead to understanding specific morphological features.
Fortunately, ghost cytometry \cite{Ota:Science:2018,Ota:arXiv2019} also showed better results of image reconstruction from raw signals on the basis of ghost imaging and compressed sensing \cite{Bioucas-DiasFigueiredo:TIP2007}. That is, we can obtain the images of the representative cells from raw signals off-line.

\section{Conclusion}
We provided a theoretical analysis of ghost features in ghost imaging and ghost cytometry.
It states that there exists a ghost feature map from an object space into a signal space that can preserve the pairwise Euclidean distances  in terms of the Frobenius norm up to a bounded relative error.
To the best of our knowledge, this work is the first step to statistically analyze and justify optical machine learning.

One direction in optical machine learning is learning structured illumination patterns from training data where first the learning process of  illumination patterns is done in computational simulation and then the learned illumination patterns are implemented in optical elements.
Since the entries in the structured illumination need to be binary, the recent advances in binarized neural networks may help with learning structured illumination patterns.
One limitation of ghost cytometry is that the illumination pattern is generated by DOEs, which is the same as existing diffractive optical neural networks.
That is, the fully hardware-implemented pattern lack flexibility in the learning process.
The spatial light modulator (SLM) can be a solution for this problem.
That is, a hybrid system comprising binarized neural networks, SLM, a single-pixel detector, and FPGA may be the next trend in optical machine learning. In this case, it is important to construct a solid theory on learning binarized neural networks constrained by SLM-based optical operations. Moreover, we need consider transfer learning and fine tuning theories from computer simulations to experiments with optical elements.

\newpage
\bibliography{reference}
\bibliographystyle{unsrt}

\newpage
\appendix
\section{Proofs}
\section{Proof of Lemma \ref{lemma:gi:exp_ineq:part_1}}
For independent random variables $Z_1$, $Z_2$, and $Z_3$ where $\bbE[Z_1]=\bbE[Z_2]=\bbE[Z_3]=0$,
we have
\begin{align}
\mathrm{Cov(Z_1Z_2,Z_2Z_3)}&=\bbE[Z_1Z_2^2Z_3]-\bbE[Z_1Z_2]\bbE[Z_2Z_3]=0,\\
\mathrm{Cov(Z_1^2,Z_1Z_2)}&=\bbE[Z_1^3Z_2]-\bbE[Z_1^2]\bbE[Z_1Z_2]=0.
\end{align}
By applying these results, Theorem \ref{theo:dependent_cov}, and
\begin{align}
(G_m(X)-q\bbS[X])^2
&=\left(\sum_{i=1}^{H}\sum_{j=1}^{W}(B_m(i,j)-q)X(i,j)\right)^2\nn\\
&=\sum_{i=1}^{H} \sum_{j=1}^{W}X(i,j)^2(B_m(i,i)-q)^2\nn\\
&\quad+2\sum_{j=1}^{W}\sum_{i=1}^{H}\sum_{k'>(j-1)H+i}^{WH}X(i,j)X(i',j')(B_m(i,j)-q) (B_m(i',j')-q),
\end{align}
we have 
\begin{align}
&\bbE\left[\exp\left(\frac{t}{q(1-q)M}\sum_{m=1}^{M} (G_m(X)-q\bbS[X])^2\right)\right]\nn\\
&=\prod_{m=1}^{M}\bbE\left[\exp\left(\frac{t}{q(1-q)M} (G_m(X)-q\bbS[X])^2\right)\right]\nn\\
&=\prod_{m=1}^{M}\prod_{i=1}^{H}\prod_{j=1}^{W}\bbE\left[\exp\left(\frac{t}{q(1-q)M}X(i,j)^2(B_m(i,j)-q)^2\right)\right]\nn\\
&~~\times  \prod_{m=1}^{M} \prod_{j=1}^{W}\prod_{i=1}^{H}\prod_{k'>(j-1)H+i}^{WH}
\bbE\left[\exp\left(\frac{2t}{q(1-q)M}X(i,j)X(i',j')(B_m(i,j)-q) (B_m(i',j')-q)\right)\right].\label{eq:exp:part_1}
\end{align}

Note that
\begin{align}
&\bbE[(B_m(i,j)-q)^2]=q(1-q),\\
&\bbE[(B_m(i,j)-q)(B_m(i',j')-q)]=0~((i,j)\not=(i',j')),\\
&\bbV[(B_m(i,j)-q)^2]\nn\\
&\quad=\bbE[(B_m(i,j)-q)^4]-\bbE[(B_m(i,j)-q)^2]^2\nn\\
&\quad=q(1-q)^4+(1-q)(-q)^4-(q(1-q))^2\nn\\
&\quad=q(1-q)((1-q)^3+q^3-q(1-q))\nn\\
&\quad=q(1-q)(1-4q+4q^2)=q(1-q)(1-2q)^2,\\
&\bbV[(B_m(i,j)-q)(B_m(i',j')-q)]~((i,j)\not=(i',j'))\nn\\
&\quad=\bbE[(B_m(i,j)-q)^2(B_m(i',j')-q)^2]-\bbE[(B_m(i,j)-q)(B_m(i',j')-q)]^2\nn\\
&\quad=\bbE[(B_m(i,j)-q)^2]\bbE[(B_m(i',j')-q)^2]=q^2(1-q)^2.
\end{align}

By using Bernstein inequality, for any $\displaystyle \frac{|t|}{M} <\frac{1}{\Lambda_q(X)}$,
\begin{align}
&\bbE\left[\exp\left(\frac{t}{q(1-q)M}X(i,j)^2(B_m(i,j)-q)^2\right)\right]\nn\\
&\quad\leq \exp\left(\frac{t}{M}X(i,j)^2+\frac{\left(\frac{t}{q(1-q)M}X(i,j)^2\right)^2\bbV[(B_m(i,j)-q)^2]}{2\left(1-\frac{|t|}{M} \Lambda_q(X)\right)}\right)\nn\\
&\quad=\exp\left(\frac{t}{M}X(i,j)^2+\frac{\left(\frac{t}{q(1-q)M}X(i,j)^2\right)^2q(1-q)(1-2q)^2}{2\left(1-\frac{|t|}{M} \Lambda_q(X)\right)}\right)
\end{align}
and
\begin{align}
&\bbE\left[\exp\left(\frac{2t}{q(1-q)M}X(i,j)X(i',j')(B_m(i,j)-q) (B_m(i',j')-q)\right)\right]\nn\\
&\quad\leq  \exp\left(\frac{\left(\frac{2t}{q(1-q)M}X(i,j)X(i',j')\right)^2\bbV[(B_m(i,j)-q)(B_m(i',j')-q)]}{2\left(1-\frac{ \Lambda_q(X)}{M}|t|\right)}\right)\nn\\
&\quad=  \exp\left(\frac{\left(\frac{2t}{q(1-q)M}X(i,j)X(i',j')\right)^2(q(1-q))^2}{2\left(1-\frac{ \Lambda_q(X)}{M}|t|\right)}\right).
\end{align}
That is, for $i'=[k'\%H]$ and $j'=\lfloor k'/H\rfloor$, 
\begin{align}
&\bbE\left[\exp\left(\frac{t}{q(1-q)M}\sum_{m=1}^{M} (G_m(X)-q\bbS[X])^2\right)\right]\nn\\
&\leq \prod_{m=1}^{M}\prod_{i=1}^{H}\prod_{j=1}^{W} \exp\left(\frac{t}{M}X(i,j)^2+\frac{\left(\frac{t\rho_1}{q(1-q)M}X(i,j)^2\right)^2q(1-q)(1-2q)^2}{2\left(1-\frac{\Lambda_q(X)}{M} |t|\right)}\right)\nn\\
&\quad\times  \prod_{m=1}^{M} \prod_{j=1}^{W}\prod_{i=1}^{H}\prod_{k'>(j-1)H+i}^{WH}
\exp\left(\frac{\left(\frac{2t}{q(1-q)M}X(i,j)X(i',j')\right)^2(q(1-q))^2}{2\left(1-\frac{\Lambda_q(X)}{M} |t|\right)}\right)\nn\\
&= \exp\left(t\|X\|_{\mathrm{F}}^2+\frac{\frac{t^2}{M}\sum_{j=1}^{W}\sum_{i=1}^{H}X(i,j)^4  \frac{(1-2q)^2}{q(1-q)}}{2\left(1-\frac{\Lambda_q(X)}{M}|t|\right)}\right)\nn\\
&\quad\times  
\exp\left(\frac{\frac{4t^2}{M}\sum_{j=1}^{W}\sum_{i=1}^{H}\sum_{k'>(j-1)H+i}^{WH}(X(i,j)X(i',j'))^2}{2\left(1-\frac{\Lambda_q(X)}{M} |t| \right)}\right)\nn\\
&= \exp\left(t\|X\|_{\mathrm{F}}^2+\frac{\Gamma_q(X)}{2M\left(1-\frac{\Lambda_q(X)}{M}|t| \right)}t^2\right).\label{eq:part1:exp_ineq}
\end{align}

Finally, we have
\begin{align}
&\bbE\left[\exp\left(\frac{t}{q(1-q)}(q\bbS[X]-\langle G(X) \rangle)^2\right)\right]\nn\\
&=\bbE\left[\exp\left(\frac{t}{q(1-q)}\left(\frac{1}{M} \sum_{m=1}^{M}G_m(X)-q\bbS[X]\right)^2\right)\right]\nn\\
&=\bbE\left[\exp\left(\frac{t}{q(1-q)M^2} \left(\sum_{m=1}^{M}\sum_{i=1}^{H}\sum_{j=1}^{W}X(i,j)(B_m(i,j)-q)\right)^2\right)\right]\nn\\
&\leq\bbE\left[\exp\left(\frac{t}{q(1-q)M^2} \sum_{m=1}^{M}\left(\sum_{i=1}^{H}\sum_{j=1}^{W}X(i,j)(B_m(i,j)-q)\right)^2\right)\right]\nn\\
&=\bbE\left[\exp\left(\frac{t}{q(1-q)M^2} \sum_{m=1}^{M}\left(G_m(X)-q\bbS[X]\right)^2\right)\right]\nn\\
&\leq \exp\left(\frac{t}{M}\|X\|_{\mathrm{F}}^2+\frac{\Gamma_q(X)}{2M^3\left(1-\frac{\Lambda_q(X)}{M^2}|t| \right)}t^2\right).
\end{align}

\section{Proof of Lemma \ref{prop:gf:linear}}
By using the linearity of $G_m$ and  $\langle G_m(X-Y) \rangle$,
\begin{align}
G_m(X-Y)
&=\sum_{i,j}B_m(i,j)(X_{i,j}-Y_{i,j})=\sum_{i,j}B_m(i,j)X_{i,j}-\sum_{i,j}B_m(i,j)Y_{i,j}\nn\\
&=G_m(X)-G_m(Y)\\
\langle G_m(X-Y) \rangle
&=\frac{1}{M}\sum_{m=1}^{M}G_m(X-Y)=\frac{1}{M}\sum_{m=1}^{M}G_m(X)-G_m(Y)\nn\\
&=\langle G_m(X) \rangle-\langle G_m(Y) \rangle,
\end{align}
we have
\begin{align}
g_m(X-Y)
&=G_m(X-Y)-\langle G_m(X-Y)=G_m(X)-G_m(Y)-( \langle G_m(X)- \langle G_m(Y) ) \rangle\nn\\
&=G_m(X)-\langle G_m(X) - (G_m(Y)-\langle G_m(Y) ).
\end{align}

\section{Proof of Theorem \ref{theorem:JL:L2}}
Note that
\begin{align}
&\sum_{m=1}^{M}(G_m(X)-\langle G(X) \rangle)^2\nn\\
&=\sum_{m=1}^{M}((G_m(X)-q\bbS[X])+(q\bbS[X]-\langle G(X) \rangle))^2\nn\\
&=\sum_{m=1}^{M}[(G_m(X)-q\bbS[X])^2+(q\bbS[X]-\langle G(X) \rangle)^2+2(G_m(X)-q\bbS[X])(q\bbS[X]-\langle G(X) \rangle)]\nn\\
&=\sum_{m=1}^{M}(G_m(X)-q\bbS[X])^2+M(q\bbS[X]-\langle G(X) \rangle)^2+2\sum_{m=1}^{M}(G_m(X)-q\bbS[X])(q\bbS[X]-\langle G(X) \rangle).
\end{align}
We now have
\begin{align}
&\sum_{m=1}^{M}(G_m(X)-q\bbS[X])(q\bbS[X]-\langle G(X) \rangle)\nn\\
&=\sum_{m=1}^{M}(G_m(X)q\bbS[X]-q\bbS[X]q\bbS[X]-G_m(X)\langle G(X) \rangle+q\bbS[X]\langle G(X) )\nn\\
&=\sum_{m=1}^{M}G_m(X)q\bbS[X]-Mq\bbS[X]q\bbS[X]-\sum_{m=1}^{M}G_m(X)\langle G(X) \rangle + Mq\bbS[X]\langle G(X) \rangle)\nn\\
&=M\langle G_(X)\rangle q\bbS[X]-Mq\bbS[X]q\bbS[X]-M\langle G(X) \rangle\langle G(X) \rangle +Mq\bbS[X]\langle G(X) \rangle)\nn\\
&=-M(q\bbS[X]q\bbS[X]+\langle G(X) \rangle\langle G(X) \rangle -2 q\bbS[X]\langle G(X) \rangle)\nn\\
&=-M(q\bbS[X]-\langle G(X) \rangle)^2.
\end{align}
Thus, we obtain
\begin{align}
\sum_{m=1}^{M}(G_m(X)-\langle G(X) \rangle)^2
=\sum_{m=1}^{M}(G_m(X)-q\bbS[X])^2-M(q\bbS[X]-\langle G(X) \rangle)^2
\end{align}
On the basis of H\"{O}lder's inequality, 
\begin{align}
&\bbE\left[\exp\left(\frac{t}{q(1-q)M}\sum_{m=1}^{M}g_m(X)^2\right)\right]\nn\\
&=\bbE\left[\exp\left(\frac{t}{q(1-q)M}\sum_{m=1}^{M}(G_m(X)-\langle G(X) \rangle)^2\right)\right]\nn\\
&=\bbE\left[\exp\left(\frac{t}{q(1-q)M}\sum_{m=1}^{M}((G_m(X)-q\bbS[X])^2-M(q\bbS[X]-\langle G(X) \rangle)^2)\right)\right]\nn\\
&=\bbE\left[\exp\left(\frac{t}{q(1-q)M}\sum_{m=1}^{M}(G_m(X)-q\bbS[X])^2\right)\exp\left(-\frac{t}{q(1-q)}(q\bbS[X]-\langle G(X) \rangle)^2\right)\right]\nn\\
&\leq\bbE\left[\exp\left(\frac{t\rho_1}{q(1-q)M}\sum_{m=1}^{M}((G_m(X)-q\bbS[X])^2\right)\right]^{\frac{1}{\rho_1}}\bbE\left[\exp\left(-\frac{t}{q(1-q)}\rho_2(q\bbS[X]-\langle G(X) \rangle)^2)\right)\right]^{\frac{1}{\rho_2}}.
\end{align}

By using Lemma \ref{lemma:gi:exp_ineq:part_1},
when $\rho_1=\rho_2=2$, 
\begin{align}
&\bbE\left[\exp\left(\frac{t}{q(1-q)M}\sum_{m=1}^{M}g_m(X)^2\right)\right]\nn\\
&\leq \exp\left(t\|X\|_{\mathrm{F}}^2+\frac{\Gamma_q(X)}{2M\left(1-\frac{\Lambda_q(X)}{M} \rho_1|t|\right)}\rho_1t^2\right)\nn\\
&\quad\times \exp\left(-\frac{t}{M}\|X\|_{\mathrm{F}}^2+\frac{\Gamma_q(X)}{2M^3\left(1-\frac{\Lambda_q(X)}{M^2}\rho_2|t| \right)}\rho_2t^2\right)\nn\\
&\leq \exp\left(t\left(1-\frac{1}{M}\right)\|X\|_{\mathrm{F}}^2+\left(1+\frac{1}{M^2}\right)\frac{\Gamma_q(X)}{M\left(1-2\frac{\Lambda_q(X)}{M} |t|\right)}t^2\right).
\end{align}
The last inequality is given by
$\left(1-2\frac{\Lambda_q(X)}{M^2}|t| \right)^{-1}<\left(1-2\frac{\Lambda_q(X)}{M}|t| \right)^{-1}$.

\noindent Thus, we have
\begin{align}
&\bbE\left[\exp\left(\frac{t}{q(1-q)M}\sum_{m=1}^{M}g_m(X)^2\right)\right]\nn\\
&\leq\exp\left(t\left(1-\frac{1}{M}\right)\|X\|_{\mathrm{F}}^2+\left(1+\frac{1}{M^2}\right)\frac{\Gamma_q(X)}{M\left(1-2\frac{\Lambda_q(X)}{M} |t|\right)}t^2\right)
\end{align}
and by using Theorem \ref{theo:bernstein_ineq} 
\begin{align}
&\bbP\left[\frac{1}{q(1-q)M}\sum_{m=1}^{M}g_m(X)^2-\left(1-\frac{1}{M}\right)\|X\|_{\mathrm{F}}^2\geq \epsilon\|X\|_{\mathrm{F}}^2\right]\nn\\
&\leq \exp\left(-\frac{(\epsilon\|X\|_{\mathrm{F}}^2)^2}{2\left(1+\frac{1}{M^2}\right)\frac{2\Gamma_q(X)}{M}+2\frac{\Lambda_q(X)}{M}\epsilon\|X\|_{\mathrm{F}}^2}\right)\nn\\
&= \exp\left(-\frac{(\epsilon\|X\|_{\mathrm{F}}^2)^2M}{2\left(\left(1+\frac{2}{M^2}\right)\Gamma_q(X)+\Lambda_q(X)\epsilon\|X\|_{\mathrm{F}}^2\right)}\right)\nn\\
&= \exp\left(-\frac{\epsilon^2M}{2\left(\left(1+\frac{2}{M^2}\right)\frac{\Gamma_q(X)}{\|X\|_{\mathrm{F}}^4}+\frac{\Lambda_q(X)}{\|X\|_{\mathrm{F}}^2}\epsilon\right)}\right).
\end{align}
In the same way, we have
\begin{align}
&\bbP\left[\frac{1}{q(1-q)M}\sum_{m=1}^{M}g_m(X)^2-\left(1+\frac{1}{M}\right)\|X\|_{\mathrm{F}}^2q(1-q)\leq \epsilon\|X\|_{\mathrm{F}}^2\right]\nn\\
&\leq \exp\left(-\frac{\epsilon^2M}{2\left(\left(1+\frac{2}{M^2}\right)\frac{\Gamma_q(X)}{\|X\|_{\mathrm{F}}^4}+\frac{\Lambda_q(X)}{\|X\|_{\mathrm{F}}^2}\epsilon\right)}\right).
\end{align}

Therefore, for every real matrix $X$, with probability at least $1-\delta$,
\begin{align}
\left(1-\frac{1}{M}-\epsilon\right)\|X\|_{\mathrm{F}}^2\leq \frac{1}{Mq(1-q)}\|g(X)\|_2^2\leq \left(1-\frac{1}{M}+\epsilon \right) \|X\|_{\mathrm{F}}^2.\label{eq:rip}
\end{align}
On the basis of the linearity of $g(X)$ (Proposition \ref{prop:gf:linear}),
substitute $X-Y$ for $X$ in Eq.\,\eqref{eq:rip}.

\section{Proof of Theorem \ref{theorem:JL:gc}}
Deriving the exponential inequality  is similar to that of ghost imaging in Eq.\eqref{eq:part1:exp_ineq}.
\begin{align}
&\sum_{m=1}^{M+W-1} (G_m(X)-q\bbS[X])^2\nn\\
&=\underbrace{\sum_{m=1}^{M}\sum_{i=1}^{H} \left(\sum_{j=1}^{W}X(i,j)^2(B(i,m)-q)^2\right)}_{\displaystyle \text{Part (A)}}\nn\\
&\quad+\underbrace{2\sum_{m=1}^{M}\sum_{i=1}^{H}\sum_{k'>(m-1)H+i}^{(m-1+W)H}\left(\sum_{j=1}^{W-(m'-m)}X(i,j)X(i',j+m'-m)\right) V_{(m-1)H+i}V_{k'}}_{\displaystyle \text{Part (B)}},
\end{align}
where $i'=[k'\%H]$ and $m'=\lfloor k'/H\rfloor$.
That is,  for any $\displaystyle \frac{t}{M} <\frac{1}{\Phi_q(X)}$,
 \begin{align}
&\bbE\left[\exp\left(\frac{t}{q(1-q)(M+W-1)}\sum_{m=1}^{M+W-1} (G_m(X)-q\bbS[X])^2\right)\right]\nn\\
&\leq\underbrace{\prod_{m=1}^{M}\prod_{i=1}^{H}\bbE\left[\exp\left(\frac{t}{q(1-q)M}\left(\sum_{j=1}^{W}X(i,j)^2(B(i,m)-q)^2\right)\right)\right]}_{\displaystyle \text{Part (A')}}\nn\\
&\quad\times\underbrace{\prod_{m=1}^{M}\prod_{i=1}^{H}\prod_{k'>(m-1)H+i}^{(m-1+W)H}\bbE\left[\exp\left(\frac{2t}{q(1-q)M}\left(\sum_{j=1}^{W-(m'-m)}X(i,j)X(i',j+m'-m)\right) V_{(m-1)H+i}V_{k'}\right)\right]}_{\displaystyle \text{Part (B')}}\nn\\
&\leq \prod_{m=1}^{M}\prod_{i=1}^{H}\exp\left(\frac{t}{M}X(i,j)^2+\frac{\left(\frac{t\rho_1}{q(1-q)M}\sum_{j=1}^{W} X(i,j)^2\right)^2\frac{(1-2q)^2}{q(1-q)}}{2\left(1-\frac{\Psi(X)}{M} |t|\right)}\right)\nn\\
&\quad\times  \prod_{m=1}^{M}\prod_{i=1}^{H}\prod_{k'>(m-1)H+i}^{(m-1+W)H}
\exp\left(\frac{\left(\frac{2t}{q(1-q)M}\left(\sum_{j=1}^{W-(m'-m)}X(i,j)X(i',j+m'-m)\right)\right)^2(q(1-q))^2}{2\left(1-\frac{\Psi(X)}{M} |t|\right)}\right)\nn\\
&= \exp\left(t\|X\|_{\mathrm{F}}^2+\frac{\frac{t^2}{M}\sum_{i=1}^{H}\left(\sum_{j=1}^{W}X(i,j)^2\right)^2\frac{(1-2q)^2}{q(1-q)} }{2\left(1-\frac{\Psi(X)}{M} |t|\right)}\right)\nn\\
&\quad\times  
\exp\left(\frac{\frac{4t^2}{M}\sum_{i=1}^{H}\sum_{k'>(m-1)H+i}^{(m-1+W)H}\left(\sum_{j=1}^{W-(m'-m)}X(i,j)X(i',j+m'-m)\right)^2}{2\left(1-\frac{\Psi(X)}{M} |t| \right)}\right)\nn\\
&\leq\exp\left(t\|X\|_{\mathrm{F}}^2+\frac{\Psi_q(X)}{2M\left(1-\frac{\Psi(X)}{M} |t| \right)}t^2\right).
\end{align}

Theorem \ref{theorem:JL:gc} holds as a consequence of Theorem \ref{theo:bernstein_ineq} and the linearity of $G(\cdot)$, i.e., $G(X-Y)=G(X)-G(Y)$.

\section{Proof of Corollary \ref{col:subExponentiality}}
By using Markov inequality, for any $\lambda>0$,
\begin{align}
\bbP[Z-\bbE[Z] \geq \epsilon]
&= \bbP[\exp(\lambda (Z-\bbE[Z])) \geq \exp(\lambda \epsilon)] \leq \frac{\bbE[\exp(\lambda (Z-\bbE[Z]))]}{\exp(\lambda \epsilon)}\nn\\
&\underset{\eqref{exp_inq:subExp}}{\leq} \exp\left(\bbV[Z]\lambda^2-\epsilon\lambda\right)
=\exp\left(\bbV[Z]\left(\lambda-\frac{\epsilon}{2\bbV[Z]}\right)^2 -\frac{\epsilon^2}{4\bbV[Z]}\right).
\end{align}
Thus, when $\lambda=\frac{\epsilon}{2\bbV[Z]}$, we have $\bbP[Z \geq \epsilon]\leq\exp\left( -\frac{\epsilon^2}{4\bbV[Z]}\right)$ for $\left|\frac{\epsilon}{2\bbV[Z]}\right| <\frac{1}{2C}$.

\end{document}